\documentclass[twocolumn,aps,pre,superscriptaddress,showpacs,floatfix]{revtex4}
\usepackage[pdfborder=0 0 0]{hyperref}
\usepackage[dvips]{graphicx}
\usepackage{amsmath}
\usepackage{color}
\usepackage{amsfonts}
\usepackage{amssymb}
\usepackage{epstopdf}
\DeclareGraphicsExtensions{.eps,.png,.pdf}
\begin{document}

\markboth{Sethia}{Sethia}
\title{Amplitude mediated \textit{chimera} states}
\author{Gautam C Sethia}
\email[e-mail: ]{gautam.sethia@gmail.com}
\affiliation{Institute for Plasma Research, Bhat, Gandhinagar 382 428, India}
\affiliation{Max-Planck-Institute for Physics of Complex Systems, 01187 Dresden, Germany}
\author{Abhijit Sen}
\affiliation{Institute for Plasma Research, Bhat, Gandhinagar 382 428, India}
\author{George L. Johnston}
\affiliation{EduTron Corp., 5 Cox Road, Winchester, MA 01890, USA}
\date{\today}
\newcommand{\beq}{\begin{equation}}
\newcommand{\eeq}{\end{equation}}
\newcommand{\beqstar}{\[}
\newcommand{\eeqstar}{\]}
\newcommand{\bea}{\begin{eqnarray}}
\newcommand{\eea}{\end{eqnarray}}
\newcommand{\beastar}{\begin{eqnarray*}}
\newcommand{\eeastar}{\end{eqnarray*}}

\begin{abstract}
We investigate the possibility of obtaining {\it chimera} state solutions of the non-local Complex Ginzburg-Landau Equation (\textit{NLCGLE}) in the strong coupling limit when it is important to retain amplitude variations. Our numerical studies reveal the existence of a variety of amplitude mediated {\it chimera} states (including stationary and  non-stationary two cluster \textit{chimera} states), that display intermittent emergence and decay of amplitude dips in their phase incoherent regions. The existence regions of the single-cluster chimera state and both types of two cluster chimera states are mapped numerically in the parameter space of $C_1$ and $C_2$ the linear and nonlinear dispersion coefficients respectively of the \textit{NLCGLE}. They represent a
new domain of dynamical behaviour in the well explored rich phase diagram of this system. The amplitude mediated {\it chimera} states may find useful applications in understanding spatio-temporal patterns found in fluid flow experiments and other strongly coupled systems. 
\end{abstract}

 
\maketitle
{\it Chimera} states, spatio-temporal patterns of co-existing coherent and incoherent behaviour in an array of coupled identical oscillators, have received a great deal of attention in recent times \cite{abrams04,abrams06,abrams08,motter10,smart12}.
 First found by Kuramoto and Battoghtokh \cite{kuramoto02} from numerical investigations of the weak coupling version of the  non-local Complex Ginzburg-Landau Equation (\textit{NLCGLE}), the {\it chimera} state has subsequently been studied for a variety of systems \cite{kawamura07,sheeba09,laing09a,laing09b,laing10,sheeba10,martens10a,martens10b,ma10,bordyugov10,sen10,lee11,olmi11,wolfrum11a,wolfrum11b,wildie12,laing12,panaggio13,pazo13,bountis13}
including two dimensional ones \cite{shima04,kuramoto06,martens10,panaggio13} and those that have time-delayed coupling \cite{sethia08} or those with a time-delayed feedback \cite{omelchenko08}.  The phase only {\it chimera} state has been suggested as a useful paradigm to represent such curious phenomenon as uni-hemispheric sleep in certain mammals and birds, where during sleep one half of their brain is quiescent while the other half remains active \cite{abrams08,rattenborg00}.  Recently the phase only chimera states have also been observed experimentally in a chemical system \cite{tinsley12} , in an opto-electronic set up \cite{hagerstrom12} under controlled laboratory settings as well as in a mechanical experiment consisting of two populations of metronomes \cite{martens13}. An experimental realization of a modified Ikeda time-delayed equation is also shown to exhibit chimera-like states \cite{larger13}.These past studies have however been confined to the weak coupling limit of the oscillator arrays where the amplitude variations have been ignored and only the dynamical behaviour of the phases have been considered. In many practical situations, such as in fluid flow representations, amplitude equations provide a more realistic description of the physical phenomena and have been widely employed to study the collective behaviour of such systems. An interesting question to ask is therefore whether spatio-temporal patterns corresponding to {\it chimera} states can exist for the strong coupling limit.  We note here that recently, multi-chimera states have been found in networks of coupled FitzHugh-Nagumo (\textit{FHN}) and Hindmarsh-Rose (\textit{HR}) neuron models \cite{omelchenko13,hizanidis13}. 
In this paper we address the question of the existence of chimera states in strong coupling limits. We report the numerical discovery of {\it chimera} solutions for the \textit{NLCGLE} equation in the regime where amplitude effects matter.  In contrast to the classical {\it chimera} states found for the phase only systems, the present ones display amplitude activity in the incoherent region of the solution in the form of intermittent emergence and decay of amplitude dips and we classify them as amplitude mediated chimeras ($AMC$s). The phases of the oscillators in the incoherent region continue to have a random distribution. These states bear a close resemblance to  the simultaneous appearance of laminar and turbulent regions in Couette flow studies\cite{barkley05,brethouwer12} and may have wider applications to other strongly coupled systems. 

\begin{figure*}
\raisebox{3.4cm} {(a)}\includegraphics[width = 0.3 \textwidth]{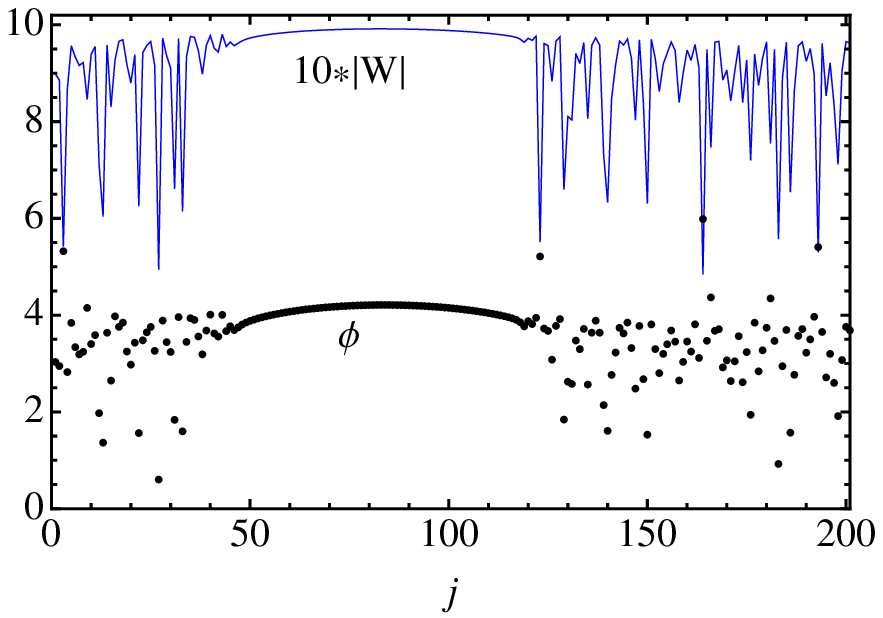}  
\raisebox{3.4cm} {(b)}\includegraphics[width = 0.3 \textwidth]{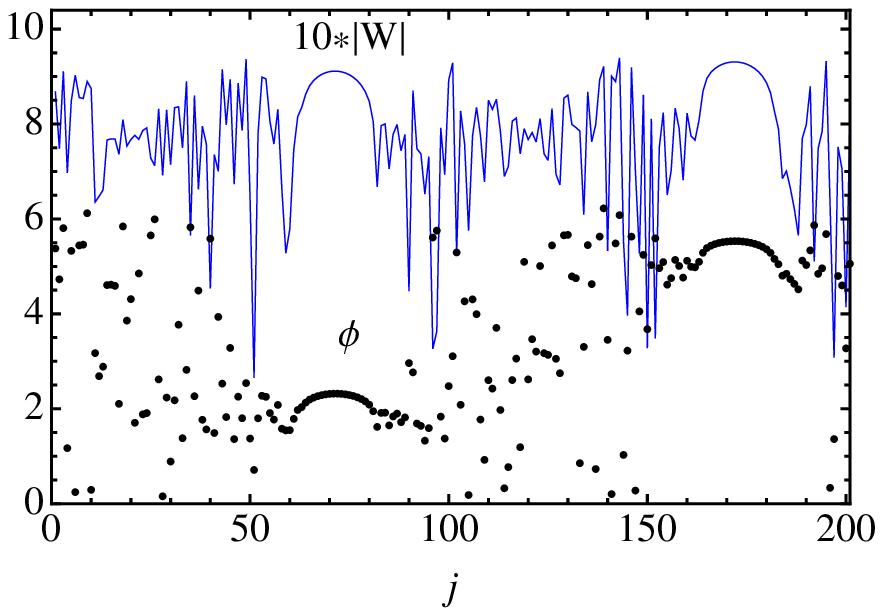} 
\raisebox{3.4cm} {(c)}\includegraphics[width = 0.3 \textwidth]{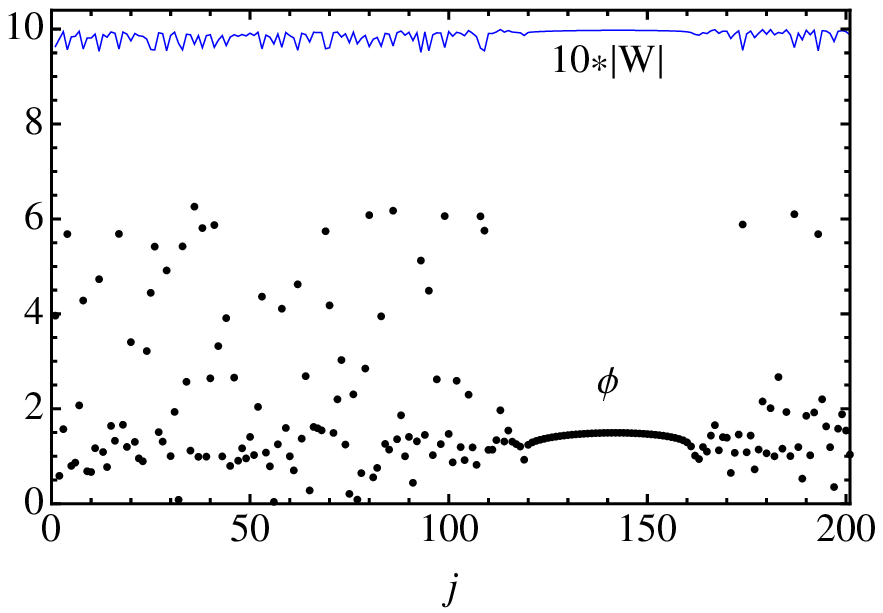} \\
\raisebox{3.4cm} {(d)}\includegraphics[width = 0.3 \textwidth]{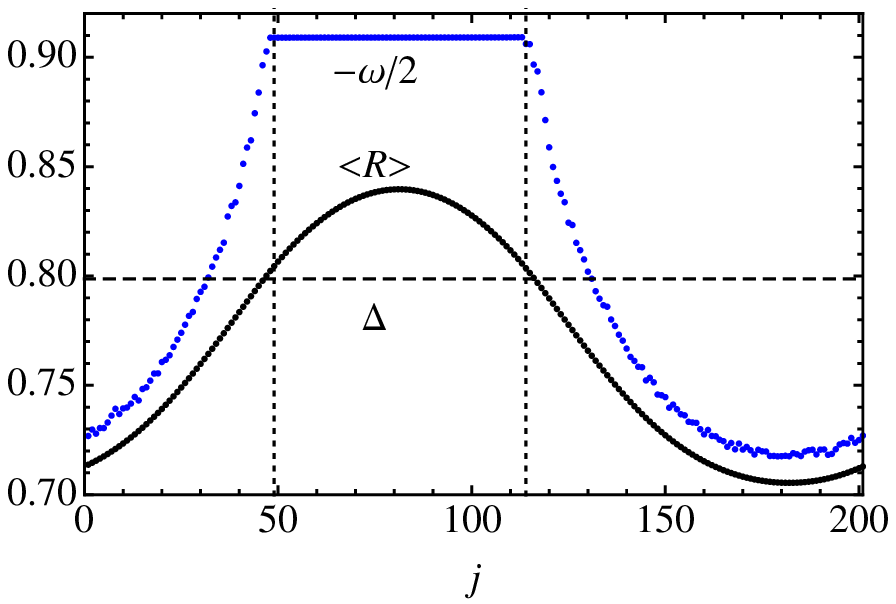}
\raisebox{3.4cm} {(e)}\includegraphics[width = 0.3 \textwidth]{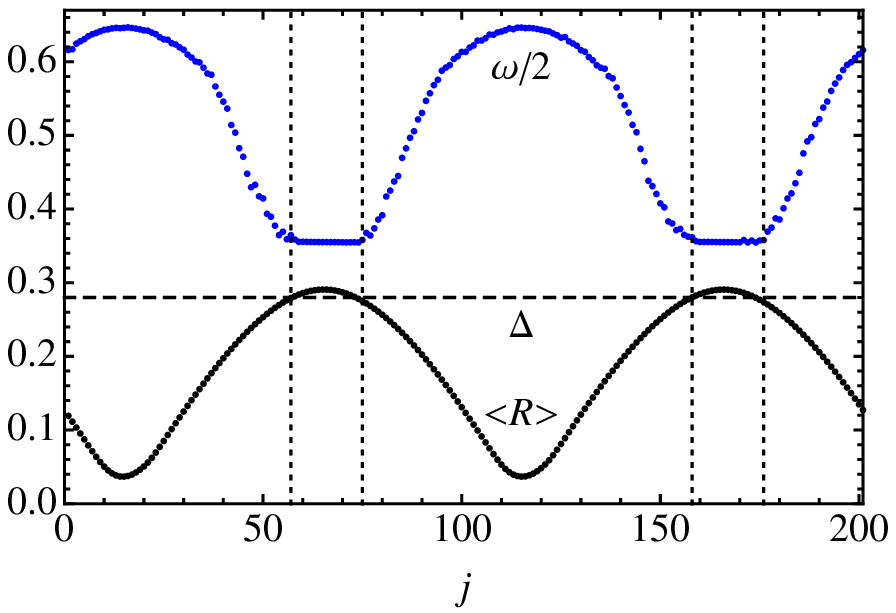} 
\raisebox{3.4cm} {(f)}\includegraphics[width = 0.3 \textwidth]{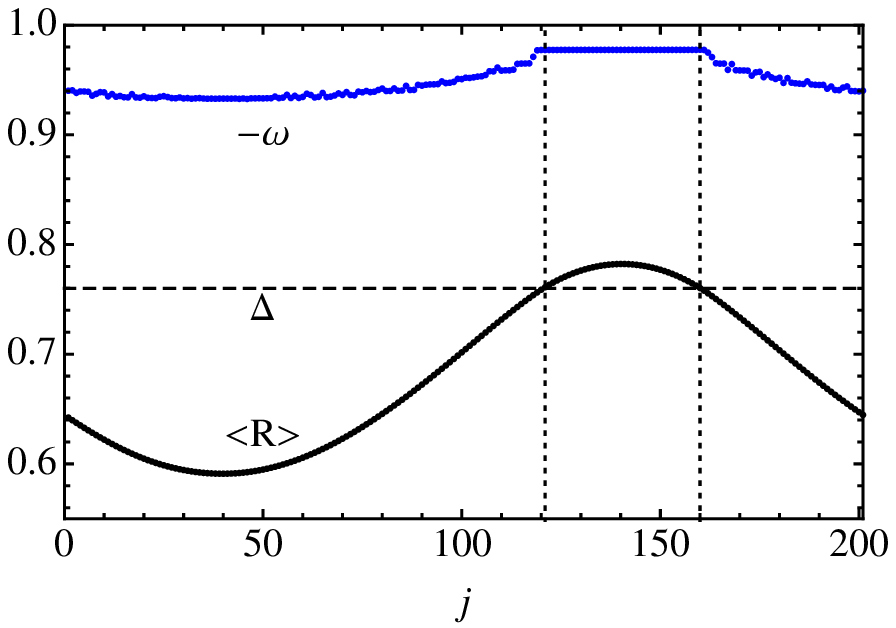} 
\caption{(Color online) Snapshots of the stationary state spatial profiles of  the phase ($\phi$) and the amplitude $|W|$ (multiplied by 10 and shown in blue, the upper curves) for $(a)$ a single cluster $AMC$ with $ K=0.40, C_1=-0.5$, $C_2=2.0$; $(b)$ a 2-cluster $AMC$ with $ K=0.40, C_1=-4.0$, $C_2=0.5$; and $(c)$ a phase-only chimera with $ K=0.05, C_1=-0.9$, $C_2=1.0$ respectively.  $\kappa$ is fixed at $2$ for all the cases. Panels $(d-f)$ show the corresponding spatial profiles of the long-time average of the order parameter amplitudes ($R$)  and the  frequencies ($\omega=<\dot{\phi}>$). The computed value of  $\Delta$ is marked with a horizontal dashed line (in black). The vertical dotted lines in the lower three panels  $(d-f)$ are drawn as a visual guidance to the coherent segments.}
\label{fig:fig1}
\end{figure*}
Our model system is the well known one dimensional \textit{NLCGLE} \cite{tanaka03} that has been extensively studied  in the past in the context of applications to various physical, chemical and biological phenomena \cite{battogtokh00,garcia-morales08,bressloff08}. 
\beq 
\frac{\partial W}{\partial t} = W-(1+i C_2)|W|^{2}W+K (1+i C_1) (\overline{W}-W)
\label{eq:nlcgle} 
\eeq 
where $W(x,t) = A(x,t) exp[i \phi(x,t)]$ is a complex field quantity with $A(x,t)$ and $\phi(x,t)$ representing the amplitude and phase respectively. $C_1$, $C_2$ and $K$ are real constants characterizing the linear and the non-linear dispersion and the coupling strength respectively.
The non-local mean field $\overline{W}(x,t)$ is given by 
\beq
\overline{W}(x,t)=\int_{-1}^{1}G(x-x') W(x-x',t) dx'
\label{eq:mean}
\eeq
where the normalized coupling kernel $G(x-x')$ has an exponentially decaying form, namely,
\beq
G(x) = \frac{\kappa }{2(1 - e^{-\kappa})}e^{-\kappa |x| }
\label{kernel}
\eeq
with $\kappa > 0$. $\kappa$ is the inverse of the coupling range and provides a measure of the non-locality of the coupling.  
The space coordinate is made dimensionless by normalizing it with the system length $L$ and hence the system size extends from $-1$ to $1$.
For $K << 1$, Eq.(\ref{eq:nlcgle}) can be reduced to a non-local evolution equation for the phase function $\phi(x,t)$ that has been the subject of several past studies for the classical {\it chimera} solutions. We have carried out extensive numerical explorations to seek {\it chimera} solutions of Eq.(\ref{eq:nlcgle}) and have 
 discovered a variety of such states over a wide range of parameter space. Broadly they consist of stationary one cluster and two cluster {\it chimera} states and also a non-stationary ({\it breather}) variety of the two cluster state.  The two coherent regions of the two cluster chimera states have opposite phases and are separated by incoherent regions. Multi-cluster phase coherent regions have also been observed before in time delayed systems \cite{sethia08} and more recently by Zhu \textit{et al.} \cite{zhu12,zhu13} in the weak coupling limit of the \textit{NLCGLE}.
 The major difference of the present solutions from their counter parts of the phase only systems is that these amplitude mediated {\it chimera} states have significant temporal variations of the amplitude in the incoherent spatial regions. These regions show intermittent emergence and decay of amplitude dips which in some cases can resemble amplitude hole (defect) solutions. 
 
 In looking for chimera states, our choice of the system parameters has been guided by earlier investigations of the \textit{NLCGLE}
including those in the weak coupling limit. Thus we have chosen two values of $K$, namely $0.05$ and $0.4$ to represent
weak and strong coupling cases respectively. The value of $\kappa$ has been chosen to be equal to $2$ so that $\kappa L =4$ which is the same as
chosen by Kuramoto and Battogtok\cite{kuramoto02} (who had $\kappa=4$ and $L=1$). The values of
$C_1$ and $C_2$ have been varied over a wide range.  
In  Fig.\ref{fig:fig1} we show a typical snapshot of the amplitude mediated one cluster and two cluster solutions in panels (a) and (b) respectively whereas in panel (c) for comparison we display a classical phase only chimera obtained in this case 
from Eq.(\ref{eq:nlcgle}) in the weak coupling limit by taking a low value of $K$. Notice that for the classical {\it chimera} state the amplitude fluctuations are negligible, justifying their neglect in the weakly coupled limit of the \textit{NLCGLE}. For the classical (phase only) chimera the set of values
$C_1 = -0.9$ and $C_2=1$  corresponds to $\alpha= \tan^{-1}(C_2-C_1) /(1+C_1 C_2) = 1.52$  where $\alpha$ is the phase-shift parameter in the weak coupling limit \cite{kuramoto02}. For the \textit{AMC}s the amplitude variations in the incoherent region are quite significant, sometimes dipping close to zero values corresponding to 
traveling \textit{hole} like solutions. Our simulations have been done with the \textit{XPPAUT} \cite{xppaut} package with $201$ discrete oscillators equally spaced on a ring. We have carefully checked our numerical results to rule out finite size effects. The nature of the \textit{AMC} is found not to change when, for example, we change
the number of oscillators from $201$ to $401$. The characteristics of the \textit{AMC }remain the same
and it does not exhibit any transient nature or tendency to collapse. The initial conditions of our simulations consist of slightly perturbed uniformly spaced phases from $0$ to $2\pi$ with unit amplitude. 

\begin{figure}
\raisebox{4.cm} {(a)}\includegraphics[width = 0.4\textwidth]{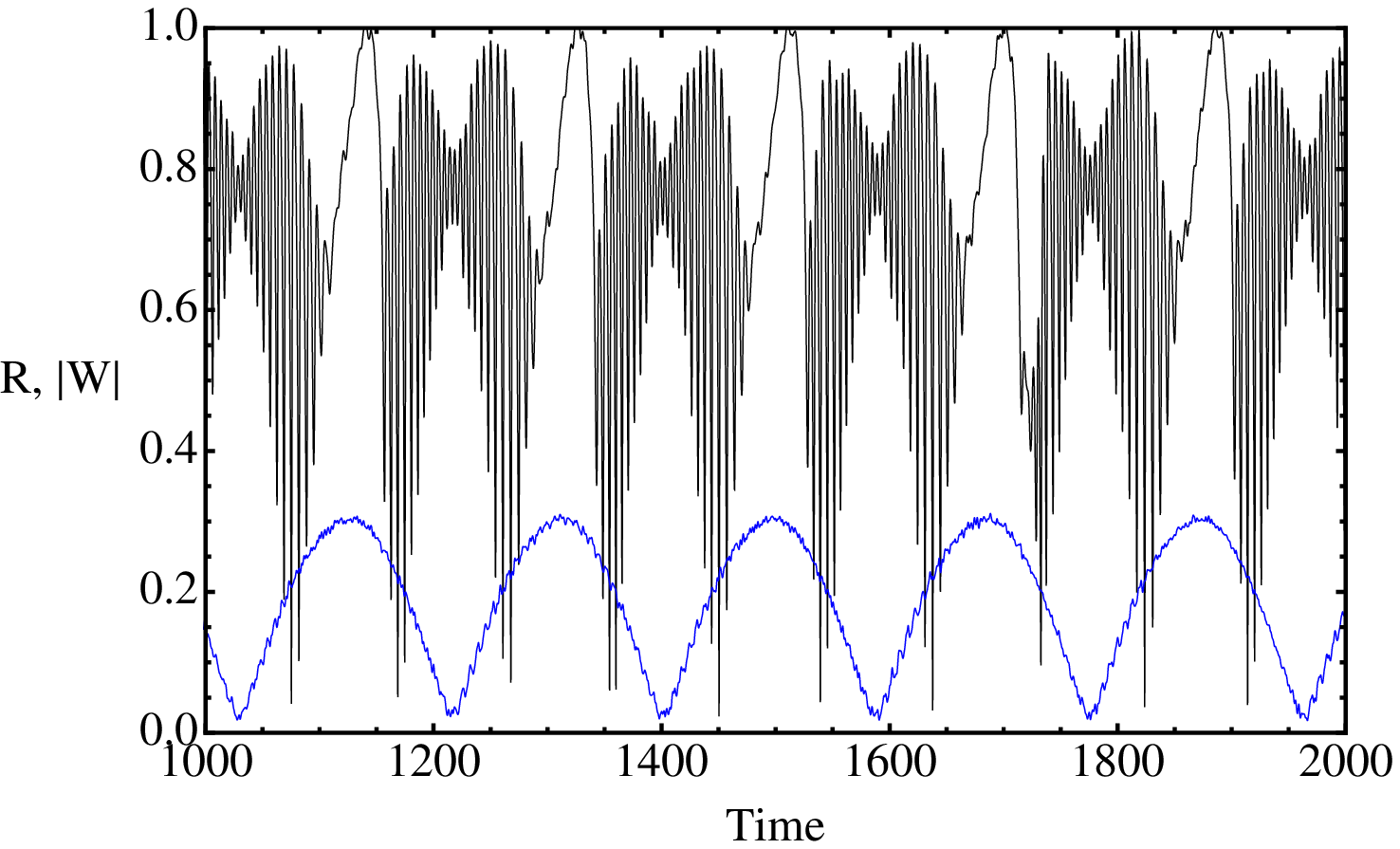} \\
\raisebox{5.2cm} {(b)}\includegraphics[width = 0.4 \textwidth]{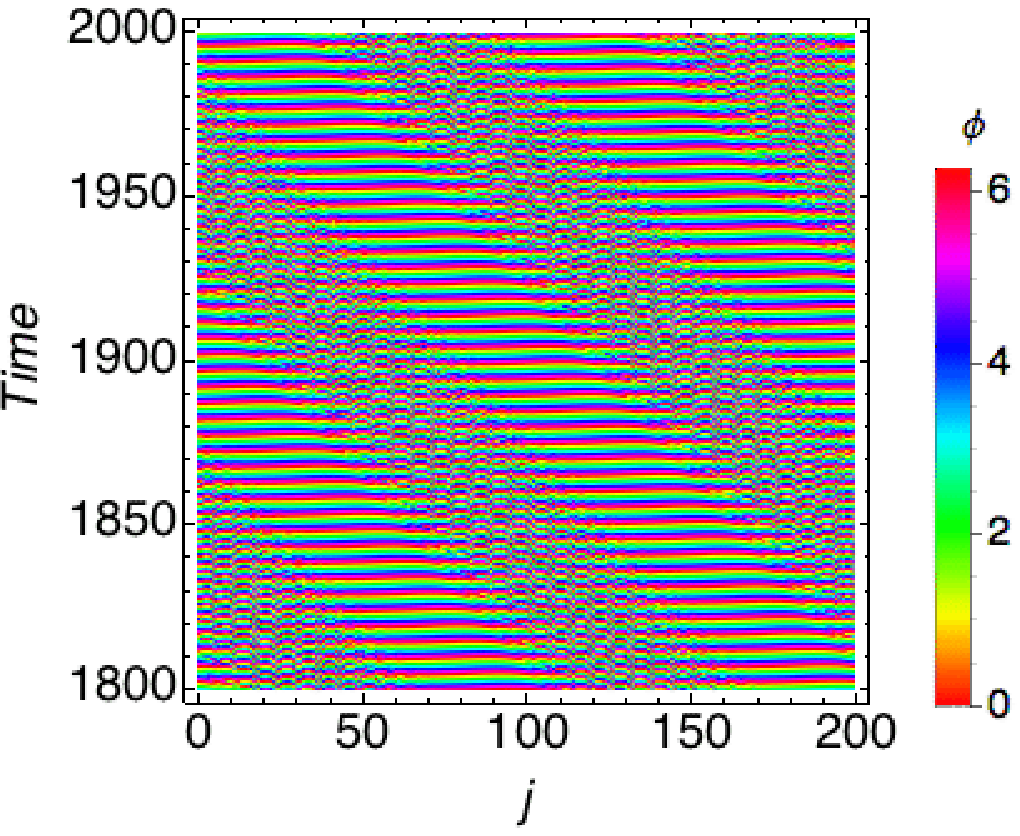} 
\caption{(Color online)  (a) The temporal patterns  of the amplitude $|W|$ of an oscillator and amplitude $R$ (the lower curve in blue) of the order parameter. (b) The spatio-temporal pattern of the phases. The parameter values for this simulation are $\kappa=2, K=0.4, C_1=-8$ and $C_2=0.95$.}
\label{fig:fig2}
\end{figure}

The stationary  patterns of these \textit{AMC}s can also be understood in terms of a complex order parameter defined as, 
\beq
R(x,t)e^{i\Theta(x,t)}=\int_{-1}^{1}G(x-x')A(x-x',t)e^{i\theta(x-x',t)}dx'
\label{eq:op}
\eeq
where $R(x,t)$ is the amplitude of the order parameter, $\Theta(x,t)$ is the mean phase and $\theta=\phi+\Omega t$ is the relative phase defined in the frame rotating with the angular
frequency $\Omega$ and amplitude $A$ of the coherent segment of the {\it chimera}.
Using Eq.(\ref{eq:op}), and separating  Eq.(\ref{eq:nlcgle}) into  its real and imaginary parts, one can get,
\bea
\frac{\partial A}{\partial t}&=&(1-K-A^2)A\nonumber\\
&&+KR\cos(\Theta-\theta)-KR C_1\sin(\Theta-\theta)\nonumber\\
A\frac{\partial \theta}{\partial t}&=&-(-\Omega+KC_1+C_2 A^2)A\nonumber\\
&&+KRC_1\cos(\Theta-\theta)+KR\sin(\Theta-\theta)
\eea 
By restricting to time stationary solutions,  we get:
\beq
\cos(\Theta-\theta)=\left [1+\frac{(1+C_1C_2)A^2-(1+C_1\Omega)}{K(1+C_1^2)}\right] \left(\frac{A}{R(x)}\right)
\label{eq:cond}
\eeq
The absolute value of the right hand side of Eq.(\ref{eq:cond}) cannot be greater than $1$ and this puts a condition on the magnitude $R$ of the order parameter, namely  ($R(x) \ge|\Delta| $) in any coherent segment in space where
\beq
\Delta=\left [1+\frac{(1+C_1C_2)A^2-(1+C_1\Omega)}{K(1+C_1^2)}\right]A
\label{eq:del}
\eeq
We obtain the amplitude $A$ and the frequency $\Omega$ of the coherent segment from the simulations and compute the value of $\Delta$ using Eq.(\ref{eq:del}). 
In Fig.\ref{fig:fig1}(d)-(f) we have plotted the time averaged profiles of $R(x)$ for the {\it chimera} states corresponding to the snapshots in Fig.\ref{fig:fig1}(a)-(c) respectively. The horizontal line in each figure marks the computed $\Delta$ value. As can be seen the results are in good agreement in that the coherent segments found in Fig.\ref{fig:fig1} correspond to regions where the condition $R>\Delta$ is satisfied. As a further check on the nature of the collective state we have also plotted the average frequency profiles  in Fig.\ref{fig:fig1} (d)-(f)
which all show the typical signature of chimera states, namely a constant frequency in the coherent region (flat profile) and a peaked profile in the incoherent region \cite{kuramoto02}.
 For the non-stationary {\it breather} state the order parameter is no longer a constant quantity but shows a periodic temporal variation. This along with amplitude $|W|$ for any one of the oscillators is shown in Fig.\ref{fig:fig2}(a) a for the 2-cluster $AMC$ state. Fig.\ref{fig:fig2}(b) shows the spatio-temporal pattern of the phase $\phi$ after the transients. Each oscillator goes through coherent and incoherent segments periodically. 
\begin{figure*}[tb]
\raisebox{5.cm} {(a)}\includegraphics[width = 0.3\textwidth]{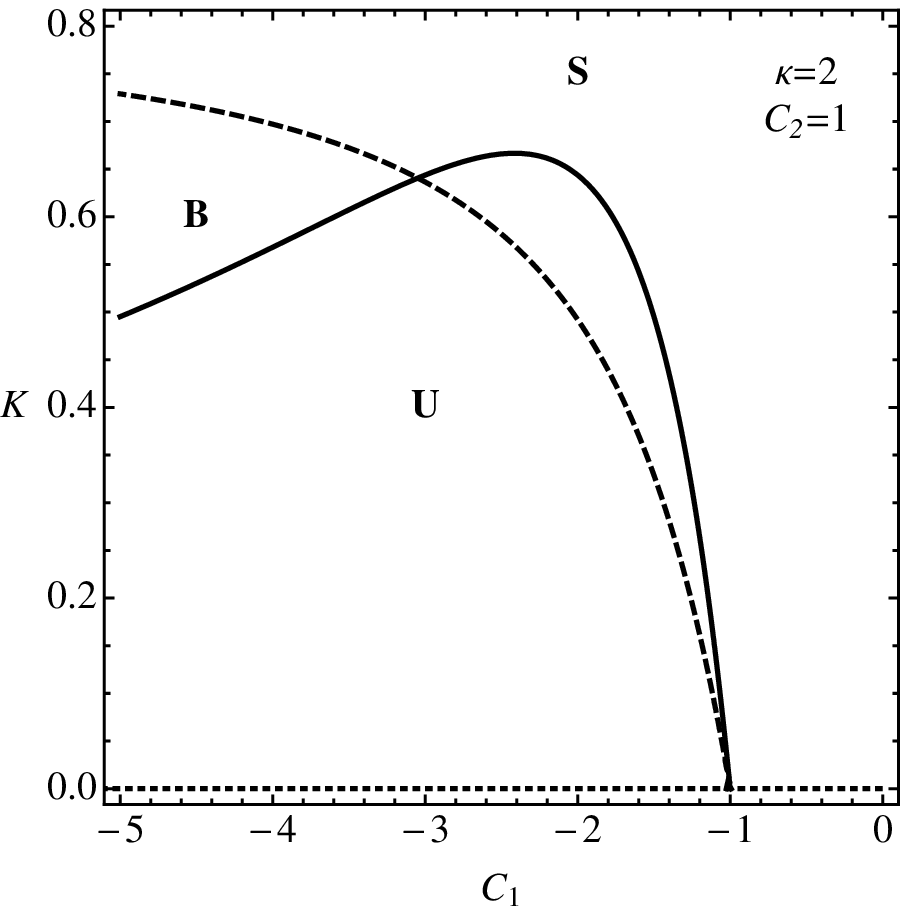} 
\raisebox{5cm} {(b)}\includegraphics[width = 0.3\textwidth]{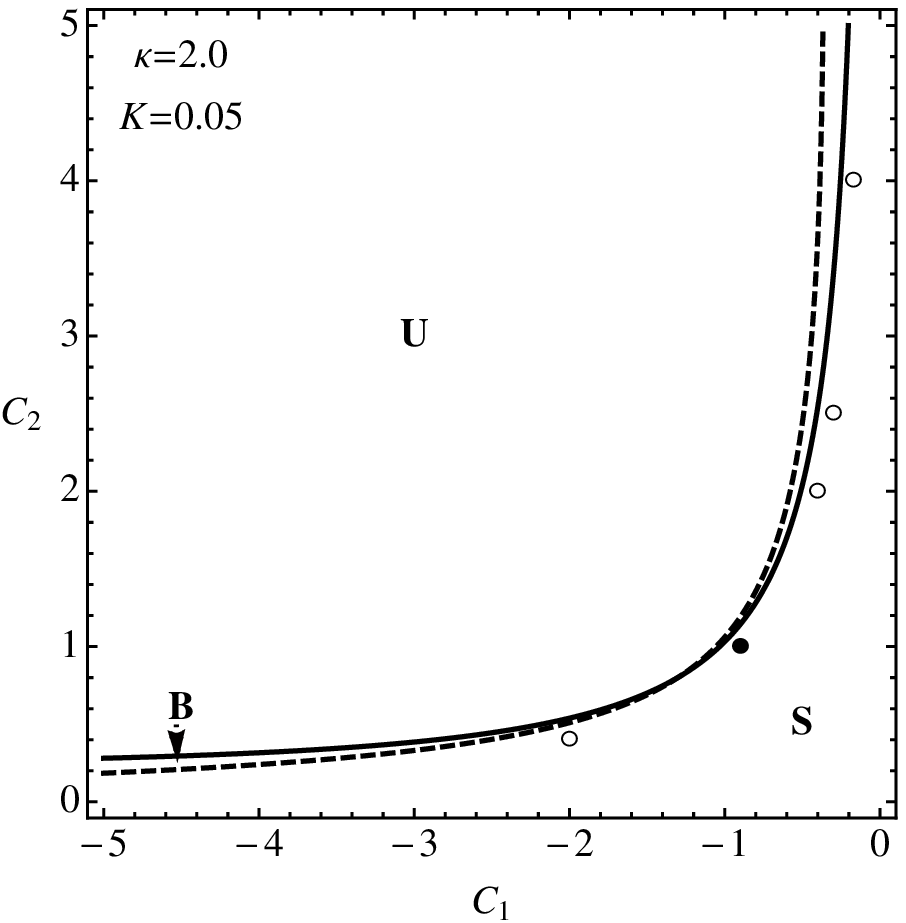} 
\raisebox{5cm} {(c)}\includegraphics[width = 0.3 \textwidth]{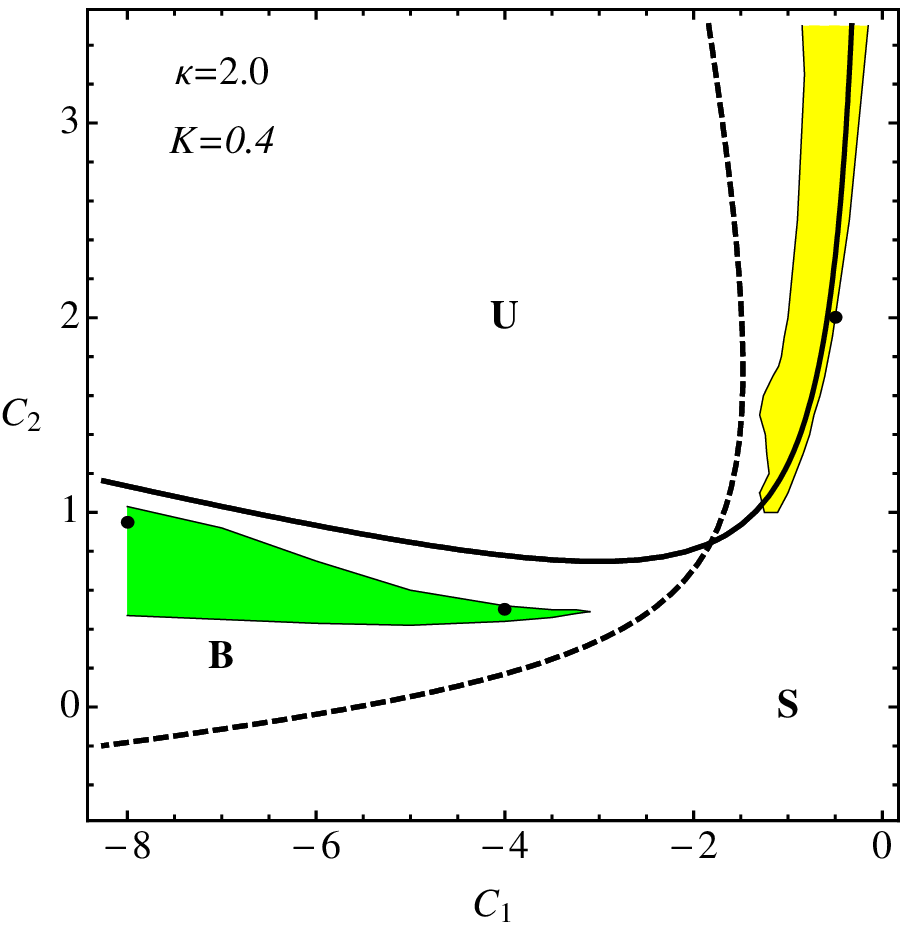} 
\caption{(Color online) $(a)$ Stability diagram of the synchronous state ($k=0$, solid curve) and the $k=1$ traveling wave (\textit{TW}) state (dashed curve) in $C_{1}-K$ plane with $\kappa=2$ and $C_{2}=1$. The region below the solid curve and marked $U$ is unstable for both the states,  region  $S$ is stable for the synchronous state and $B$ is a bistable region. $(b)$ and $(c)$ Stability diagrams similar to $(a)$ but in the phase plane of $C_{1}-C_{2}$. $K$ is chosen to be $0.05$ for $(b)$ and $0.40$ for $(c)$. The open circle symbols in (b) mark the phase only chimeras
that are found in the weak coupling limit. The filled circle marks the chimera state shown in Fig.~\ref{fig:fig1}(c). In (c) the  yellow colored region (upper  shaded region) shows the existence domain of
the single cluster \textit{AMC} and the green region (lower shaded region in bistable domain $B$) that of the two cluster stationary and breather \textit{AMC}s. The three filled circles mark the positions of the \textit{AMC}s displayed in Fig.~\ref{fig:fig1}(a) and (b)
and Fig.~\ref{fig:fig2}. }
\label{fig:fig3}
\end{figure*}

To get a perspective of the existence regions of the $AMC$s with respect to other collective states of the \textit{NLCGLE}, we have next carried out a linear stability analysis of plane wave solutions of Eq.(\ref{eq:nlcgle}), that are of the form,
$W_k^0(x,t) = a_{k} e^{i(\pi kx-\omega_{k} t )} $
and that satisfy the dispersion relation,
\beq
\omega_k=C_1 (1 - a_k^2)+C_2 a_k^2
\label{disp}
\eeq
with $a_k^2=1-K'$, $K'=K(1-I_{\kappa,k})$ and 
$$I_{\kappa,k}=\int_{-1}^{1}G(x') e^{i\pi k x'} dx'$$
Perturbing these equilibrium solutions by writing $W_k=[1+u(x,t)]W_k^0$, where $u(x,t) = \sum_{n}{u_{n}(t) e^{i\pi n x}}$ and substituting in the linearized form of Eq.(\ref{eq:nlcgle}), we can get a variational equation for $u_{n}$,
\bea
\frac{\partial u_n}{\partial t}&=&[1+i\omega_k-2(1+iC_2)a_k^2+ K(1+iC_1)\nonumber\\
&& (I_{\kappa,n+k}-1)]u_n-(1+iC_2)a_k^2\bar{u}_n
\label{eq:un}
\eea
Taking $u_n(t) \sim e^{\lambda t}$, Eq.({\ref{eq:un})  and its complex conjugate  yield a $2 \times 2$ matrix $M$, whose eigenvalues are determined from the following quadratic characteristic equation \cite{Kuramoto84book},
\beq
|M - \lambda I| \equiv \lambda^2+(r_1+ir_2)\lambda+p_1+ip_2=0
\label{eq:eigen}
\eeq
where $I$ is the $2 \times 2$ unit matrix and $r_1=-(a+e), r_2=-(b+f), p_1=-bf+ae-c^2-d^2$ and $p_2=af+be$, which in turn are expressed in terms of system parameters:
\begin{eqnarray}
a&=&1-2 a_k^2+K I_{\kappa,n+k}-K\nonumber\\
b&=&\omega_k-2C_2 a_k^2+K C_1(I_{\kappa,n+k}-1)\nonumber\\
c&=&-a_k^2\nonumber\\
d&=&-C_2a_k^2\nonumber\\
e&=&1-2a_k^2+K I_{\kappa,n-k}-K\nonumber\\
f&=&-\omega_k+2C_2 a_k^2-K C_1(I_{\kappa,n-k}-1).
\end{eqnarray}
 Eq.(\ref{eq:eigen}) determines the eigenvalues $\lambda$ for
a perturbation with a wave number $n$. Setting the real part of $\lambda$ to be zero gives us a condition for marginal stability of the form $\Phi(\kappa,k,n)=0$, 
where
\beq
\Phi(\kappa,k,n)=-p_2^2+r_1 r_2 p_2+r_1^2 p_1
\label{eq:K}
\eeq
From our numerical analysis we find that the lowest mode number perturbation ($n=1$) is the first one to get destabilized and 
therefore determines the marginal stability curve. We fix $n=1$ for our further stability analysis. For the uniform ($k=0$) state, we are able to get a simple analytic form for the marginal stability curve, namely,
\beq
\Phi(\kappa,0,1)=1+C_1C_2+\frac{K(1+C_1^2)}{2}(1-\gamma)=0
\label{eq:Phi}
\eeq
where $\gamma  \equiv I_{\kappa,1} =\frac{\kappa^2 \coth(\frac{\kappa}{2})}{\pi^2+\kappa^2}$  accounts for the non-locality in the system.

The stability condition $\Phi > 0$ reduces to  the well known Benjamin-Feir-Newell criterion $1+C_1C_2>0$ for $\gamma \to 1$ corresponding to large $\kappa$ i.e. local or diffusive coupling 
and in the global limit ($\gamma \to 0$ for small $\kappa$ ),  to that reported in earlier works \cite{hakim92,nakagawa94}.
Fig.\ref{fig:fig3}(a) shows the stability diagram 
for the uniform state ($k=0$) as well as for the $k=1$ traveling wave state in the $C_1-K$ phase space where  $C_2$ is fixed at $1$  and $\kappa$ at $2$. The phase only models are valid near the dotted line at $K=0$. Similarly Fig.\ref{fig:fig3}(b),(c) show the stability diagrams in $C_1-C_2$ phase space for two different values of $K$ but the same value of $\kappa$. The location of 
a few {\it chimera} states are marked by different point symbols (filled and open circles) on these stability diagrams. The open circles in Fig.~\ref{fig:fig3}(b) represent phase only chimera states that are found in the weak coupling limit. The filled circle marks the chimera state shown in Fig.~\ref{fig:fig1}(c). It is seen that the phase only chimera states co-exist with the
stable uniform traveling wave state ($k=0$) as has been noted earlier \cite{abrams06}. }
The \textit{AMC}s on the other hand can exist in both the stable and unstable region of the $k=0$ state. To determine the existence domain of the \textit{AMC}s we have carried out a systematic and extensive numerical exploration in the $C_1\;-\;C_2$ phase space.  Our results are shown in Fig.~\ref{fig:fig3}(c) where the existence domains are marked in color.  Single cluster \textit{AMC}s are found in the region marked yellow (upper shaded region) and the two cluster and breather \textit{AMC}s exist
in the region marked in green (lower shaded region in bistable domain $B$). These domains  thus mark a new dynamical region for the \textit{NLCGLE} representing an additional 
collective excitation state of the system.

In conclusion, we have studied the \textit{NLCGLE} system in the strong coupling limit and found a new class of {\it chimera} states where the incoherent regions display significant amplitude
fluctuations. These amplitude mediated {\it chimeras} can be of the stationary kind (with a single or two cluster configuration of coherent regions)  or have an oscillatory nature. Our detailed numerical investigation have also marked out the existence regions of these
hybrid states in the reference frame of the stability diagram of the uniform state and the $k=1$ traveling wave state of the \textit{NLCGLE}. These states not only complement the previously found phase only {\it chimera} states but also extend the applicability domain of such hybrid states to physical systems that are better represented by full blown amplitude equations such as the \textit{NLCGLE}. Some systems that come to mind in this context are fluid flow simulations/experiments where the simultaneous appearance of laminar and turbulent regions have been observed \cite{barkley05} and neuronal networks
displaying \textit{bump} states where a subset of neurons fire in synchrony while others fire incoherently \cite{laing01}. The discovery of these novel states also opens up a number of interesting future areas of investigation including a study of their stability, delineating their linkages to other coherent solutions of the \textit{NLCGLE} such as traveling waves and  \textit{holes} and exploring their existence for other forms of the coupling kernel.

GCS acknowledges the support of \textit{MPI-PKS}, Dresden, Germany, where part of the work was carried out.


\end{document}